# A Sum-Product Model as a
# Physical Basis for Shadow Fading

Jari Salo

**Abstract**

Shadow fading (slow fading) effects play a central role in mobile communication system design and analysis. Experimental evidence indicates that shadow fading exhibits log-normal power distribution almost universally, and yet it is still not well understood what causes this. In this paper, we propose a versatile sum-product signal model as a physical basis for shadow fading. Simulation results imply that the proposed model results in log-normally distributed local mean power regardless of the distributions of the interactions in the radio channel, and hence it is capable of explaining the log-normality in a wide variety of propagation scenarios. The sum-product model also includes as its special cases the conventional product model as well as the recently proposed sum model, and improves upon these by: a) being applicable in both global and local distance scales; b) being more plausible from physical point of view; c) providing better goodness-of-fit to log-normal distribution than either of these models.

## I. INTRODUCTION

In physical sciences, a theory of an observable phenomenon is developed by comparing theoretical predictions to measurements, and, possibly, modifying the theory if disagreement is found. In wireless radio channel research, one attempts to deterministically or stochastically

This work was supported by the Academy of Finland and National Technology Agency of Finland. The author was with the SMARAD Centre of Excellence, Radio Laboratory, Helsinki University of Technology. He is now with the ECE Ltd.

predict the behavior of radio waves in real-world environments by comparing channel models to measurements. If the model disagrees with experimental data, it should be revised.

The purpose of this paper is to propose an improved model for the so-called shadow fading phenomenon observed in mobile radio channel measurements. Shadow fading (slow fading, shadowing) is loosely defined as the fluctuation in the received power averaged over a small area, typically having a diameter of $10 - 40$ wavelength in outdoor environment [1], [2]. It plays a central role in various communication system design tasks, including coverage prediction for network planning and analysis of system-level radio network algorithms. In a large number of measurements conducted worldwide, the local mean power has been found to be well approximated by the log-normal distribution over a wide range of propagation environments, carrier frequencies, and transmitter-receiver distances [1]–[5]. Yet, no generally accepted explanation for shadow fading or its log-normality has been presented. Considering the great variety of propagation scenarios exhibiting log-normal power variation it seems clear that the central limit theorem, in one form or another, masks away the intricacies of the highly complex underlying propagation physics. Otherwise, it is difficult to explain why the log-normal statistics occur in environments where the dominating propagation mechanisms are different; for example, in urban and suburban macro cells [6], [7] and urban micro cells [8].

The conventional *product model* assumes that the radio signal propagating from the transmitter to the receiver experiences a cascade of random attenuations. From the central limit theorem it follows that the logarithm of the random product is approximately normally distributed [1], [3]. While this model appears physically plausible, experimental observations disagree with it to some extent. First, results from extensive computer simulations [9], [10] imply that the number of random attenuations needs to be quite large for the convergence to log-normal distribution to take place, e.g. $30 - 100$. While such a large number of interactions[1] is possible, the resulting path loss would be very high. Furthermore, measurements in urban areas indicate that most of the signal power is transported by propagation modes that experience much less than $10$ channel interactions [11]. Second, the product model implies that path loss increases exponentially with distance $d$, and that the standard deviation of shadow fading is also distance-dependent [12]–[15]. However, majority of measurements suggest a $\propto d^{\alpha}$ path loss law (typically with $\alpha = 3 \ldots 7$), and that the standard deviation of shadow fading is approximately independent of the receiver–transmitter separation [4], [5]. Hence, even if the product model is true, it seems that the number of measurable interactions is limited, so that e.g. exponential path loss difficult to detect in practice[2]. Third, an interesting measurement result with two receive antennas [18] indicates that when the patterns of the antennas are not identical (i.e., the antennas see different multipath environment), the correlation coefficient between the shadow fading processes observed at the antenna outputs was clearly less than unity. The result contradicts the product model which implicitly assumes that all multipaths experience the same multiplicative attenuative process. The observation that the variation of the local mean power is at least partly related to multipath propagation (instead of 'shadowing' by obstacles) is also supported by [19]–[21].

---

[1] In this paper, by 'interaction' we mean either reflection, diffraction, or scattering.

The *sum model* proposed in [14], [19] is based on the idea that the amplitudes of the planewaves impinging at the moving receiver change slowly in time/space with respect to small-scale fading, giving rise to slow variation in the local mean power. In a line-of-sight microcell this was shown to result in Nakagami-$m$ or log-normal distributed slow fading [19]. The idea was generalized to general environments in [14], where the variation of the power of the plane waves was modelled stochastically, and it was argued based on the central limit theorem and a simple linearization approach that the resulting local mean power may be perceived as log-normal, regardless of the power distributions of the individual plane waves. Unfortunately, the applicability of the sum model in [14] is limited to a small area, called "extended local area", and it cannot explain the measured $5 - 12$ dB path loss standard deviations in a larger, global area.

It is evident that neither the product model nor the sum model can alone explain all aspects of the shadow fading phenomenon, and its log-normality in particular. The contribution of this paper is to introduce a unified *sum-product model* that has the advantage of being more plausible from physical propagation point of view than the product model, as well as being applicable in a larger geographical area than the sum model. It is shown that, in addition to providing a more realistic physical basis, the shadow fading distribution induced by the sum-product model gives better fit to the log-normal distribution than the product model. Moreover, unlike with the sum model [14], the standard deviation of shadow fading can assume large values often encountered in measurements (e.g. $8-12$ dB), since the justification of log-normality is not based on a linear approximation of the logarithm function, which only holds if the deviation of shadow fading relative to its mean is small.

---

[2]Some evidence to indicate that path loss is exponential has been reported in [16], while measurement results in [17] suggest that standard deviation of shadow fading is distance dependent.

As in [14], we adopt a stochastic channel modelling approach in the sense that we do not specify the underlying physical propagation mechanisms (diffraction, reflection, scattering). Instead, the channel interactions are modelled probabilistically. This approach has the advantage that the results hold more generally, rather than for any given propagation environment.

The paper is organized as follows. In Section II we introduce the sum-product model, and in Section III it is shown how the product and the sum models are special cases of it. Simulation set-up is detailed in Section IV, while the simulation results confirming the applicability of the sum-product model are given V. Section VI concludes the paper.

*Notation:* A scalar, vector, and matrix are denoted by $a$, $\mathbf{a}$, and $\mathbf{A}$, respectively, while $\mathbf{A}^T$ and $\mathbf{A}^H$ stand for the transpose and conjugate transpose of $\mathbf{A}$. The expected value of a random variable $X$ is denoted by $\mathrm{E}[X]$. A set $\{x_1, x_2, \ldots, x_N\}$ is also denoted with $\{x_n\}_{n=1}^N$, or just $\{x_n\}$ when $N$ is obvious from the context.

## II. SUM-PRODUCT MODEL FOR RADIO PROPAGATION

As the first step of this section we explain the fully determistic point-to-point version of the sum-product model, where all channel interactions are assumed constant (transmitter, receiver and channel are not moving). In the second phase, we let the receiver move within a small local area, such that the phases of the received plane waves vary in a random fashion, and we then examine the signal power averaged over small-scale fading. In the third step, the receiver is allowed to move in a large, global area, and all channel interactions are assumed random.

## A. Point-to-point response

We introduce the sum-product model by means of a simple example. Consider the point-to-point radio propagation scenario shown in Fig. 1. Crosses in the figure represent interactions experienced by the propagating plane waves (or, "rays"). The interactions in the radio channel can be either scattering, reflection, or diffraction. The rays with complex responses $\{b_1, b_2\}$ first interact with the first layer of interacting objects, and then continue onwards to interact with the second layer, and so on. At the receiver antenna output, the propagating rays are weighted by $\{a_1, a_2\}$ and summed to yield a complex scalar received signal $y$. Some remarks on the model:

- It is assumed that $\{a_n\}$, $\{b_m\}$, and $\{s_{k,ij}\}$ include the directional responses of the interacting objects at both ends of the connecting edge, including the free space path loss in between them.

- It is assumed that all channel interactions have directional responses. To give an example, consider a channel interaction (e.g. wall reflection) excited by a plane wave from a direction $\theta_1$. We then assume that the response of the channel interaction towards another direction $\theta_2$ can be written as a product of two directional responses $s_{k-1}(\theta_1)s_k(\theta_2)$.

- *Example 1 (see Fig. 1)*: the complex number $b_1$ captures the complex responses of the transmitter antenna to the direction of the channel interaction '$A$', radio channel between antenna and '$A$', and the channel interaction '$A$' towards the direction of the transmitter.

- *Example 2 (see Fig. 1)*: the complex number $\{s_{1,21}\}$ captures the complex response of the channel interaction '$A$' towards the direction of channel interaction '$B$', channel response between interactions '$A$' and '$B$', and response of channel interaction '$B$' towards channel interaction '$A$'.

This received signal at a fixed transmitter and receiver position can be written as

$$
\begin{aligned}
y \;=\; & a_1[s_{2,11}(s_{1,11}b_1 + s_{1,12}b_2) + s_{2,12}(s_{1,21}b_1 + s_{1,22}b_2)] \\
& + a_2[s_{2,21}(s_{1,11}b_1 + s_{1,12}b_2) + s_{2,22}(s_{1,21}b_1 + s_{1,22}b_2)],
\end{aligned}
$$

or in matrix form

$$y = \mathbf{a}^T \mathbf{S}_2 \mathbf{S}_1 \mathbf{b} \,, \tag{1}$$

where $\mathbf{a} = [a_1\ a_2]^T$, $\mathbf{b} = [b_1\ b_2]^T$, and

$$\mathbf{S}_k = \begin{bmatrix} s_{k,11} & s_{k,12} \\ s_{k,21} & s_{k,22} \end{bmatrix} \,, \quad k = \{1,2\} \,. \tag{2}$$

All scalars $\{a_n\}$, $\{b_m\}$, $\{s_{k,ij}\}$ are assumed complex valued. For now, we assume the propagation environment can be modelled as a network of channel interactions, as shown in Fig. 1, where direct connectivity exists only between interactions of neighboring "layers". For example, the interactions at the transmitter end in Fig. 1 are connected to the interactions at the receiver end via the two intermediate ones. Hence the graph is not fully connected, which, interpreted in terms of radio propagation, means that the path loss between certain channel interactions is so large that it can neglected. In the sequel, we shall call a vertical group of interactions a "layer of interactions". Hence, in Fig. 1, there are three layers of interactions, with two interactions in each. A matrix $\mathbf{S}_k$ characterizing the interactions between two neighboring layers is called a "coupling matrix".

More generally, we assume the signal model depicted in Fig. 2, which suggests that the received signal can be written as a bilinear form

$$y = \mathbf{a}^T \underbrace{\mathbf{S}\mathbf{b}}_{\mathbf{c}} \tag{3}$$

$$= \mathbf{a}^T \mathbf{c} \,, \tag{4}$$

where

$$\mathbf{a} = [a_1 \ \cdots \ a_N]^T$$

$$\mathbf{b} = [b_1 \ \cdots \ b_M]^T$$

$$\mathbf{c} = [c_1 \ \cdots \ c_N]^T$$

$$\mathbf{S} = \mathbf{S}_K \mathbf{S}_{K-1} \cdots \mathbf{S}_1 \,. \tag{5}$$

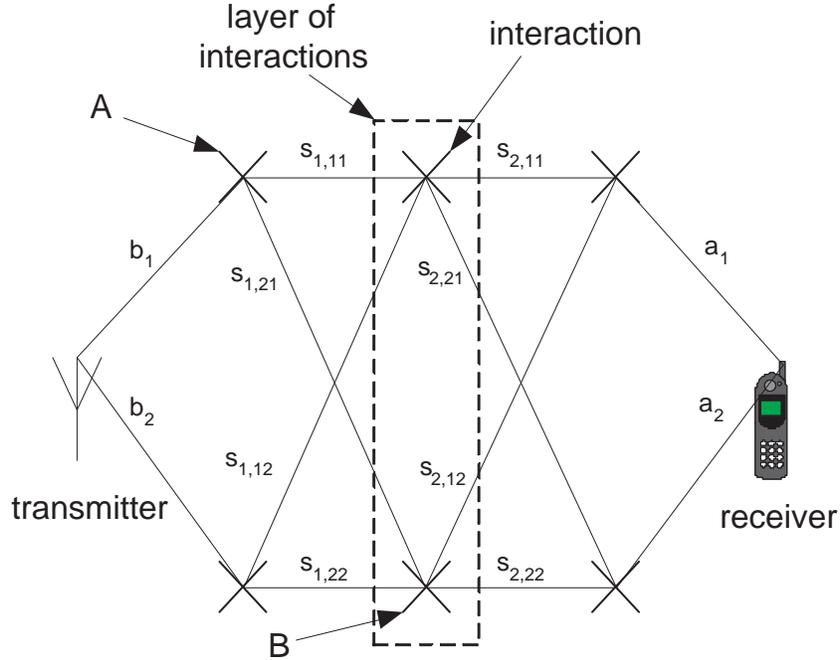

Fig. 1.   Example of the sum-product signal model with $K = M = N = 2$ (three layers of channel interactions).

The composite coupling matrix $\mathbf{S}$ characterizes how the transmitted plane waves are coupled to the received ones. In general, $\{\mathbf{S}_k\}_{k=1}^K$ need not be square. By allowing rectangular matrices, one can vary the number of interactions within a layer. Empty interaction corresponds to a column of zeros in $\mathbf{S}_k$. Of course, for (3) to make sense it is required that $\mathbf{S}_K$ has $N$ rows and that $\mathbf{S}_1$ has $M$ columns, so that $\mathbf{S}$ is $N \times M$. By imposing structure on the coupling matrices the bilinear form (3) lends itself for modelling a great variety of radio propagation scenarios; some examples will be given in Section III.

### B. Signal power averaged over small-scale fading

While the model in (3) may be applicable to wider range of purposes, in this paper we use it to study the distribution of received signal power averaged over small-scale fading. The local mean power is obtained by averaging over small-scale fading, or, equivalently, averaging over the phase of the received signal $y$. Towards this end, suppose now that the receiver in Fig. 2 is moving. To average over the small-scale fading we assume that only the phases of the entries of

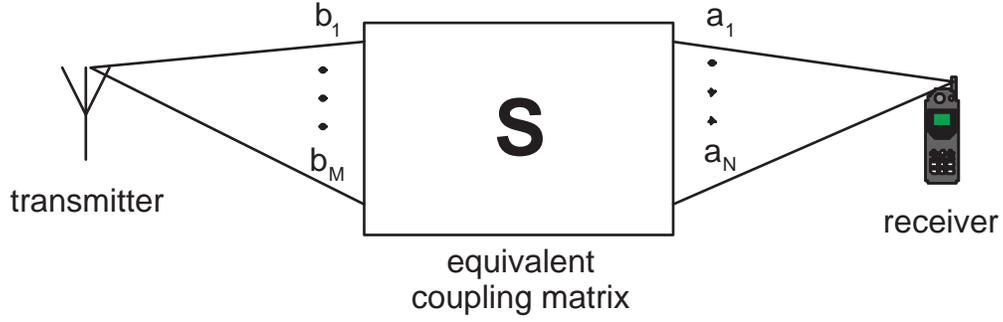

Fig. 2. Signal model. $\mathbf{S}$ is the $N \times M$ composite coupling matrix describing the mapping between the transmitted and received plane waves.

$\mathbf{a}$ are changing independently and uniformly over $[0, 2\pi)$. All other quantities, including $\{|a_n|\}$, $\{b_m\}$ and $\{s_{k,ij}\}$, are assumed constant within the area over which the average is taken[3]. In other words, the small-scale fluctuation of $y$ is assumed to be entirely due to variation in the phases of $\{a_n\}$. If the aforementioned assumptions hold, the local mean power is obtained by taking the expected value over the phases of $\mathbf{a}$:

$$
\begin{aligned}
P &= \mathrm{E}[|y|^2] \\
&= \mathrm{E}[|\mathbf{a}^T \mathbf{S} \mathbf{b}|^2] \\
&= \mathrm{E}[\mathbf{b}^H \mathbf{S}^H \mathbf{a} \mathbf{a}^H \mathbf{S} \mathbf{b}] \\
&= \mathbf{b}^H \mathbf{S}^H \mathrm{E}[\mathbf{a} \mathbf{a}^H] \mathbf{S} \mathbf{b} \\
&= \mathbf{b}^H \mathbf{S}^H \boldsymbol{\Gamma}_\mathbf{a} \mathbf{S} \mathbf{b},
\end{aligned}
\tag{6}
$$

where $\mathbf{S}$ was defined in (5), and $\boldsymbol{\Gamma}_\mathbf{a} = \mathrm{E}[\mathbf{a} \mathbf{a}^H]$ is an $N \times N$ diagonal matrix with $\{|a_n|^2\}_{n=1}^N$ on the diagonal. Alternatively, we may write (6) as

[3]In practice, one would approximate the theoretical mean power with e.g. sliding mean over measured power of the small-scale fading signal. In measurement analysis, the averaging length is conventionally chosen as about $10 - 40$ wavelengths in outdoor radio propagation environments [1].

$$P = \sum_{n=1}^{N} |a_n|^2 |c_n|^2, \tag{7}$$

where $\{c_n\}$ denote the entries of $\mathbf{c}$ [see (4)]. This expression has the same form as the one given for local area power in [14], although it has been derived from the viewpoint of the more general signal model considered in this paper.

## C. Variability of local area power ($P$) – shadow fading

In (7), it is assumed that $\{|a_n|\}$ and $\{c_n\}$ are constant, and hence $P$ is constant for a given averaging area, also called local area [14], [22]. Consider now an experiment where the power is measured over a large number of local areas, as shown in Fig. 3. Assuming that the number of rays at the transmitter ($M$) and receiver ($N$) end as well as the number of "layers" ($K$) in the signal model (3) are constant over all measured local areas, i.e., for all indices $q$ in Fig. 3, we obtain a set of local area powers: $\{P_q\}_{q=1}^{Q}$. By further assuming that $\{a_n\}$, $\{b_m\}$, $\{s_{k,ij}\}$ for all $n, m, k, i, j$ can be modelled as random variables, we can treat $\{P_q\}$ as realizations of a random variable with an unknown distribution. From the above assumptions, it follows that the average local area power, which can be interpreted as average path loss, is constant. This can be seen by taking the expected value of (7), where $\{|a_n|\}$ and $\{c_n\}$ are by assumption stationary.

In this paper, we are interested in the distribution of the local area power, $P$. In Section V we provide simulation results that imply that the sum-product model (6) results in log-normally distributed local area power, regardless of the distributions of the individual components of the model. Fig. 4 gives a preview of this interesting result. The empirical cdf of $10 \log_{10}(P)$ in Fig. 4 is from a Monte Carlo simulation where $\{|a_n|\}_{n=1}^{10}$, $\{|b_m|\}_{m=1}^{10}$, and $\{|s_{k,ij}|\}_{k=1}^{40}$ are from uniform and R distribution, to be defined in Section IV. Normal distribution is plotted for reference with dotted line. It can be seen that the empirical distribution of $10 \log_{10}(P)$ is very close to normal, i.e., $P$ is log-normally distributed.

Note that by setting $N = M = 1$ and assuming that the coupling matrices are scalars, (6) collapses to a product of $K + 2$ scalar random variables, which is known to converge in distribution to log-normal under mild conditions as $K \to \infty$. Remarkably, the simulation results in Fig. 4 and Section V suggest that the central limit theorem for the logarithm of a product of random positive scalars generalizes to the matrix quadratic form in (6); this is itself an interesting observation that requires further investigation.

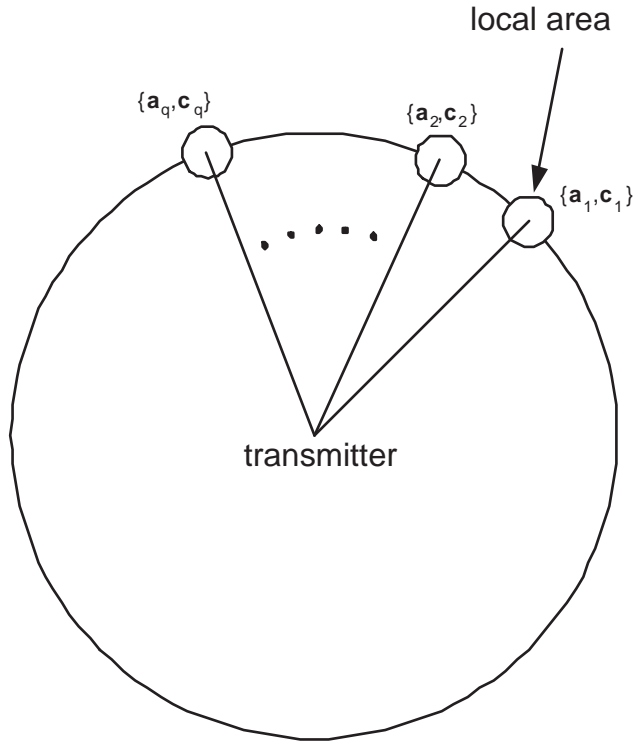

Fig. 3. Sampling of mean power over several local areas each having a random set of parameters $\{\mathbf{a}_q\}_{q=1}^Q$, $\{\mathbf{c}_q\}_{q=1}^Q$. In Section II-B, only $\{\mathbf{a}_q\}_{q=1}^Q$ are assumed random, while in Section II-C also $\{\mathbf{c}_q\}_{q=1}^Q$ are random.

## III. SPECIAL CASES OF THE GENERAL SIGNAL MODEL

Before simulation results, we briefly illustrate how the general sum-product model yields as its special cases the classical multiplicative shadow fading model and the recently proposed additive model.

### A. Multiplicative global shadow fading model

In the classical multiplicative shadow fading, all plane waves experience a cascade of random multiplicative attenuations. This can be modelled by setting[4] $\mathbf{S}_k = s_k \mathbf{I}_N$ in (6). In this special

---

[4]We assume for simplicity that all $\mathbf{S}_k$ are square.

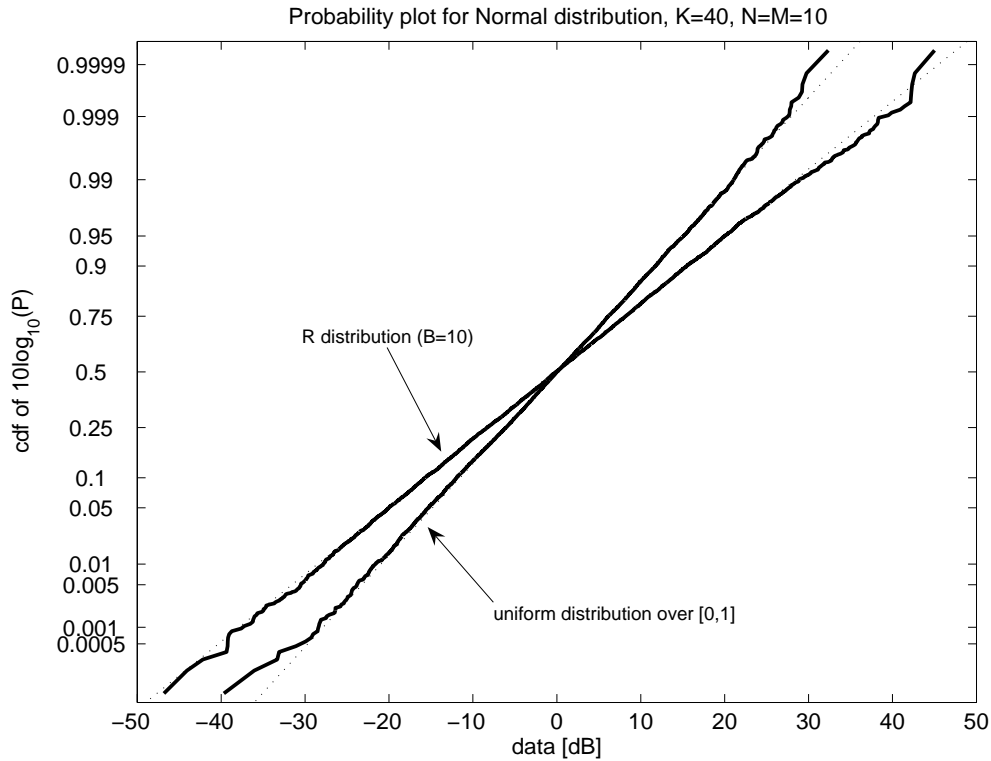

Fig. 4. Example of empirical distribution of $10 \log_{10}(P)$ with $P$ given in (6). The amplitudes of the complex entries of $\mathbf{a}$, $\mathbf{b}$, and $\{\mathbf{S}_k\}_{k=1}^{40}$ are independently drawn from R and uniform distribution over the unit interval, and the phases are uniform over $[0, 2\pi)$. Normal distribution is shown (dotted line) for reference. The R distribution is defined in Table I, Section IV.

case, (6) reduces to

$$P = \left( \sum_{n=1}^{N} |a_n|^2 |b_n|^2 \right) \cdot \prod_{k=1}^{K} |s_k|^2. \tag{8}$$

This is a product of $K + 1$ random factors, whose distribution is well-known to converge to log-normal for large $K$.

### B. Additive local shadow fading model

In [14], an additive model was proposed as a physical basis for shadow fading with an area that consists of a several local areas. From the viewpoint of the present framework, this "extended local area" is on one hand assumed large enough so that $\{|a_n|^2\}$ are randomly time-variant, and on the other hand small enough so that $\mathbf{c}$, i.e., the intermediate propagation process, can be assumed constant. If these conditions hold, $\{|c_n|^2\}$ are constant in (7), and shadow fading is only due to variation of the amplitudes of the plane waves at the receiver end. It was shown in [14] that the resulting distribution of $P$ may be perceived as log-normal if the standard deviation of $P$ relative to its mean is not too large, or if $\{|a_n|^2\}_{n=1}^N$ themselves are log-normal and $N$ is small.

## C. Other special cases

By suitable selection of the coupling matrices $\{\mathbf{S}_k\}$, it is possible to model a variety of propagation scenarios. We provide three examples.

*1) Line-of-sight:* In line-of-sight propagation, the direct ray experiences only free space loss, while other rays interact with the environment. This scenario can be modelled, for example, by setting

$$\mathbf{S}_k = \begin{bmatrix} \sqrt[k]{\mathrm{pl}} & \mathbf{0}_{1 \times (N-1)} \\ \mathbf{0}_{(N-1) \times 1} & \mathbf{S}_{k,\mathrm{nlos}} \end{bmatrix}.$$

In the above pl denotes the free space path loss, $\mathbf{0}_{n \times m}$ an $n \times m$ matrix of zeros, and $\mathbf{S}_{k,\mathrm{nlos}}$ the $k$th coupling matrix of the non-line-of-sight rays. We have assumed for simplicity that all coupling matrices are square.

*2) Keyhole propagation:* It has been shown theoretically that a keyhole degrades the information capacity of a MIMO communication system [23]. Although in this paper we focus on the single-input single-output case, we remark that the signal model (3) could also be extended to MIMO case. To model a keyhole, one would simply set one of the coupling matrices to a row vector.

*3) Cluster-based propagation:* Consider the case where the plane waves propagate in $L$ clusters so that each cluster of rays is completely decoupled from the other clusters. This scenario is realized by letting $\{\mathbf{S}_k\}$ be block diagonal, i.e.,

$$\mathbf{S}_k = \begin{bmatrix} \mathbf{S}_{k,1} & \mathbf{0} & \mathbf{0} \\ \vdots & \ddots & \vdots \\ \mathbf{0} & \mathbf{0} & \mathbf{S}_{k,L} \end{bmatrix}, \tag{9}$$

where $\{\mathbf{S}_{k,l}\}_{l=1}^L$ denotes the $k$th coupling matrix of the $l$th cluster. Further simplification results by assuming that all rays within a cluster undergo the same multiplicative attenuation, which amounts to letting $\mathbf{S}_{k,l} = s_{k,l}\mathbf{I}$, where $\mathbf{I}$ is an identity matrix.

## IV. Simulation procedure

In order to examine the distribution of shadow fading, we can let $\mathbf{b}$, $\{\mathbf{S}_k\}$, and $\mathbf{a}$ (also called "components" in the sequel) in the quadratic form (6) be random variables. Unfortunately, we have been unable to find in statistics literature any analytical results or limit theorems on the distribution of the quadratic form. Therefore, in the sequel we shall resort to Monte Carlo simulations in order to study its distribution empirically. The empirical findings, reported in Section V, indicate that the distribution of the quadratic form converges to log-normal, regardless of the distributions of its components.

### A. General assumptions

The following assumptions are made to restrict the number of free parameters in the simulations. Computer experiments indicate, however, that the conclusions in Section V hold also under more general conditions.

- We set $M = N$ and assume that the coupling matrices $\{\mathbf{S}_k\}_{k=1}^K$ are all square matrices.
- It is assumed that phases of $\{a_n\}$, $\{b_n\}$, and $\{s_{k,ij}\}$ are independent and uniformly distributed over $[0, 2\pi)$ for all $n, k, i, j$. It is also assumed that the phases are distributed independently of the corresponding amplitudes $\{|a_n|\}$, $\{|b_n|\}$, and $\{|s_{k,ij}|\}$.
- It is assumed that the amplitudes $\{|a_n|\}$, $\{|b_n|\}$, and $\{|s_{k,ij}|\}$ are identically and independently distributed for all $n, k, i, j$.

In summary, the tunable parameters in the Monte Carlo simulation are: the number of plane waves ($N$), the number of coupling matrices ($K$), and the amplitude distribution of the components.

### B. Distribution of the components

To our knowledge, there are no published measurement studies available on the distributions of the attenuations of the channel interactions. To keep the amplitude distributions physically reasonable, the support of the random variable should be restricted to unit interval $[0, 1]$, since the radio channel is a passive component and hence the energy of the propagating signal cannot increase in interactions. Assuming a positive random variable $X$ defined over $[0, \infty]$, the transformation $Y = (1 + X)^{-1}$, for example, results in another random variable $Y$ defined over the unit interval. For purposes of simulations, we shall assume that $X$ is either Rayleigh, with density

$$f_X(x) = \frac{x}{B^2} \exp\left(-\frac{x^2}{2B^2}\right),\tag{10}$$

or log-normal with density

$$f_X(x) = \frac{1}{x\sigma\sqrt{2\pi}} \exp\left(-\frac{(\ln x - \mu)^2}{2\sigma^2}\right).\tag{11}$$

The corresponding distributions of $Y$ will be called R and L distributions, for short. In addition, we shall employ the beta distribution, which for $A = B = 1$ reduces to the uniform distribution. The amplitude distributions are summarized in Table I, where the same notation as with MATLAB's random number generators has been adopted. The distributions of the attenuations (in dB scale) are illustrated for selected values of parameters in Fig. 5.

To further keep the simulation model physical, one should normalize the columns of each coupling matrix by $\sqrt{N}$, or

$$\mathbf{S}'_k = \frac{\mathbf{S}_k}{\sqrt{N}},\tag{12}$$

where the entries of $\mathbf{S}_k$ are from a distribution defined over the unit interval. This normalization ensures that the propagating energy is not amplified in channel interactions. However, it is easy to see by plugging (12) into (6) that this normalization only shifts the mean of the distribution of $\ln P$ but does not affect its shape. As we are primarily interested in the shape of the distribution of the random variable $\ln P$, we choose to ignore the power normalization (12) in the simulations. Furthermore, for easier visual comparison all distributions shown will also centered so that the mean value is $0$ dB.

TABLE I

AMPLITUDE DISTRIBUTIONS USED IN THE MONTE CARLO SIMULATIONS

| name | definition$^\dagger$ | parameters |
|------|------------|------------|
| beta | pdf: $\frac{1}{B(A,B)} y^{A-1}(1-y)^{B-1}$ | $\{A, B\}$ |
| R | $Y = \frac{1}{1+X}$, $X \sim$ Rayleigh | $B$ |
| L | $Y = \frac{1}{1+X}$, $X \sim$ log-normal | $\{\mu, \sigma\}$ |

$^\dagger$ $B(A, B)$ denotes the beta function.

## C. Quantifying goodness-of-fit

As in [9], to quantify the distance of the empirical cdf of $\ln P$ from the normal distribution, we utilize the Kolmogorov-Smirnov (K-S) test statistic [24]. The test statistic is defined as the largest absolute difference between the empirical cdf (from simulation) and the normal distribution. Note that, again as in [9], our main interest is not to test the hypothesis that the observed data comes from the log-normal distribution[5], but rather to compare the relative rates of convergence with different parameters and component distributions. This will allow us to make some conclusions of how well the sum-product model can explain the shadow fading phenomenon.

## V. SIMULATION RESULTS

The purpose of the numerical experiments in this section is to answer the following questions:

- How does the relative convergence speed of the sum-product model and the conventional product model depend on the number of interaction layers ($K$)? What are the standard deviations of the resulting distributions, and how sensitive they are to the value of $K$?
- What is the impact of the number of plane waves ($N$)?
- What is the impact of the distribution of the attenuations of the channel interactions?

In all cases, the empirical cdf and the corresponding K-S test statistic has been computed from a Monte Carlo simulation with $10^5$ realizations of the random variable $P$ [cf. (6)].

---

[5]For any nontrivial values of the parameters $N, M, K$ and any distributions of the entries of $\{\mathbf{S}_k\}, \mathbf{a}, \mathbf{b}$ in (6) it is highly unlikely that the true distribution (not known in analytic form) is *exactly* log-normal. Therefore, testing hypothesis that $\ln P$ is log-normal does not make sense since for large sample sizes the test will anyway fail with high probability. We conjecture, however, that as $K \to \infty$, $\ln P$ (suitably normalized) converges in distribution to normal.

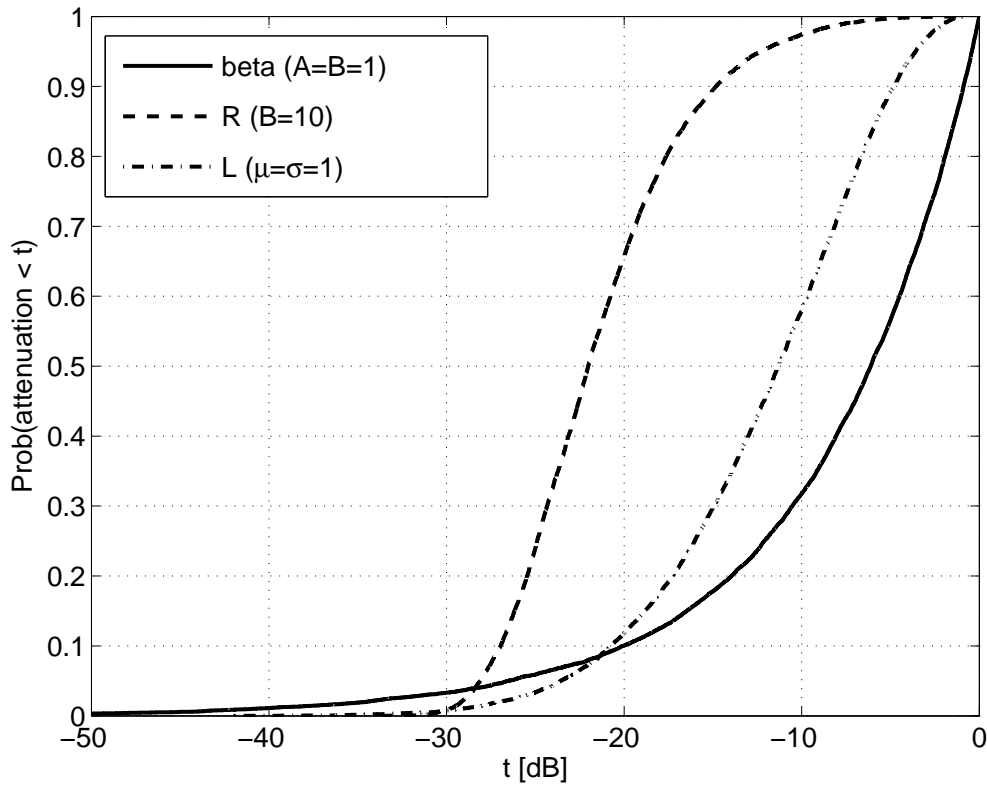

Fig. 5.   Example of cdfs.

### A. Sum-product model versus product model

The main focus of this paper is to study whether shadow fading could result from the sum-product signal model, and whether the results predicted by it are more plausible than those obtained from the 'text book' product model. To shed some light on these issues, we plot the value of K-S test statistics for increasing number of multiplicative layers, $K$. The test is repeated both for the product model in (8) and the sum-product model in (6). Fig. 6 shows the results for the case where the amplitudes of the entries of $\{\mathbf{S}_k\}$, $\mathbf{a}$, and $\mathbf{c}$ are independently generated from beta, R and L distributions (see Table I) with parameters $A = B = 1$ (uniform), $B = 10$, and $\mu = \sigma = 1$, respectively. The number of rays is $N = M = 10$.

From Fig. 6 it can be seen that for a given value of $K$, the empirical power distribution generated by the sum-product model is much more similar to the log-normal than that of the conventional product model. As implied by extensive numerical studies [9], [10], the convergence of the product model requires a large number of factors (e.g. $> 30$) to converge to log-normal distribution. On the other hand, the convergence of the sum-product model is quite fast for all tested distributions, and the fit is also considerably better.

When comparing models one should, in addition to the shape of the signal power distribution, also consider the mean path loss and the magnitude of its standard deviation. Unfortunately, it is not straightforward to compare models in terms of absolute path loss without incorporating some propagation physics into the model. This, however, would make simulations hopelessly complicated[6], and therefore in this paper we limit our focus to the stochastic approach. Nonetheless, it is of interest to compare the standard deviations of the distributions in Fig. 6 for different values of $K$, see Table II. From the table, it can be seen that the standard deviation with the product model increases quickly with $K$. With the uniformly distributed channel interactions, the deviation is 55 dB for $K = 40$, whereas for the sum-product model it is only about 9 dB. In general, with the sum-product model the standard deviation is less sensitive to $K$; for $K$ increasing from 5 to 20 the value of the deviation ranges from 3.8 to 8.9 dB; these values are well within the range of typical values encountered in measurements. From radio propagation viewpoint, the value of $K$ cannot be large, since this would result in very low received signal power owing to the multiple layers of interactions. In this sense, the relative insensitivity of the sum-product model to the value $K$ can be related to the empirical observation that shadow fading standard deviation is relatively independent of the receiver-transmitter distance [3]. On the other hand, the standard deviation of the product model is quite sensitive to the number of attenuations; little or no empirical evidence of such behavior has been observed in measurements.

### B. Impact of number of planewaves

Fig. 8 illustrates the impact of the number of planewaves ($N$) to the distribution of the sum-product model. Changing $N$ has only a minor effect on the distribution of the product model. This is because in (8) $N$ affects only one term in the product, whereas in the sum-product model changing $N$ changes also the dimensions of the coupling matrices (since we assume for

---

[6]Numerical experiments with the simple product model incorporating over-the-rooftop propagation effects was considered in [10]. Similar experiments with the sum-product model would be even more complicated.

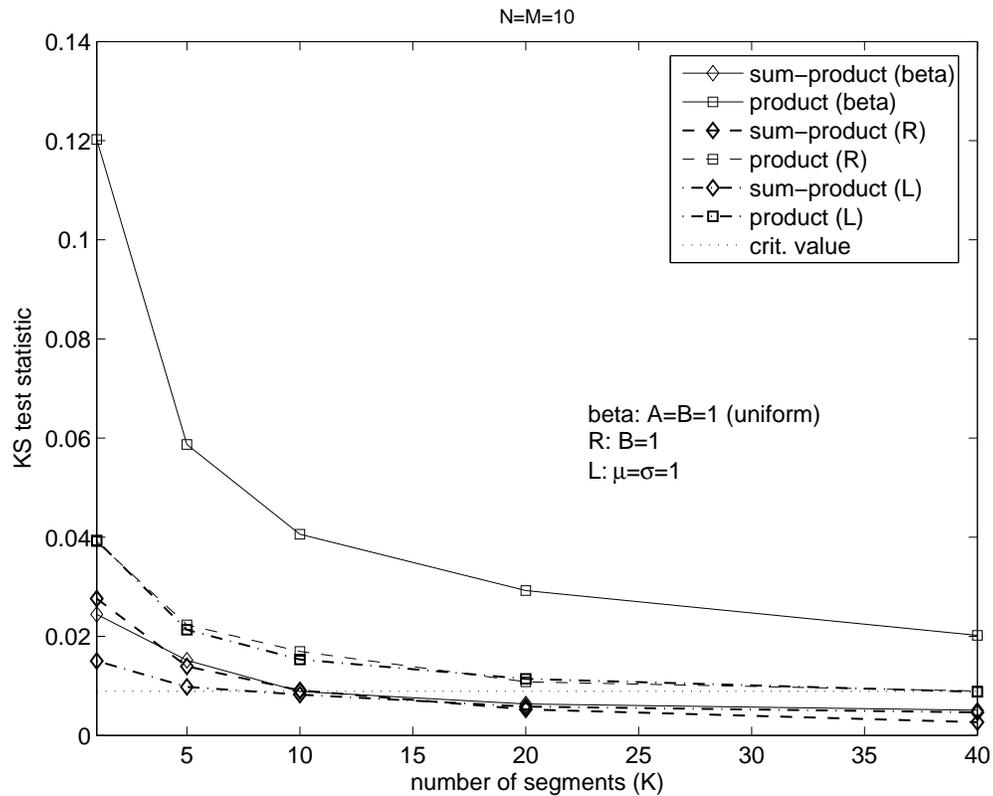

Fig. 6. Sum-product model versus product model. Small value of the test statistics indicates good fit. Critical value of the Kolmogorov-Smirnov test for 95% significance and $10^3$ samples is shown for reference. Beta: A=B=1. R: B=10. L: $\mu = \sigma = 1$.

TABLE II

STANDARD DEVIATIONS FOR FIG. 6

| model/pdf | std for $K = 1/5/10/20/40$ [dB] |
|-----------|--------------------------------|
| sumprod/beta | 2.7/3.8/4.9/6.6/9.1 |
| sumprod/R | 4.2/5.6/6.9/8.9/12.0 |
| sumprod/L | 3.1/4.2/5.3/7.0/9.5 |
| prod/beta | 9.0/19.5/27.5/38.8/55.1 |
| prod/R | 6.1/11.4/15.6/21.7/30.6 |
| prod/L | 6.7/14.1/19.6/27.7/39.0 |

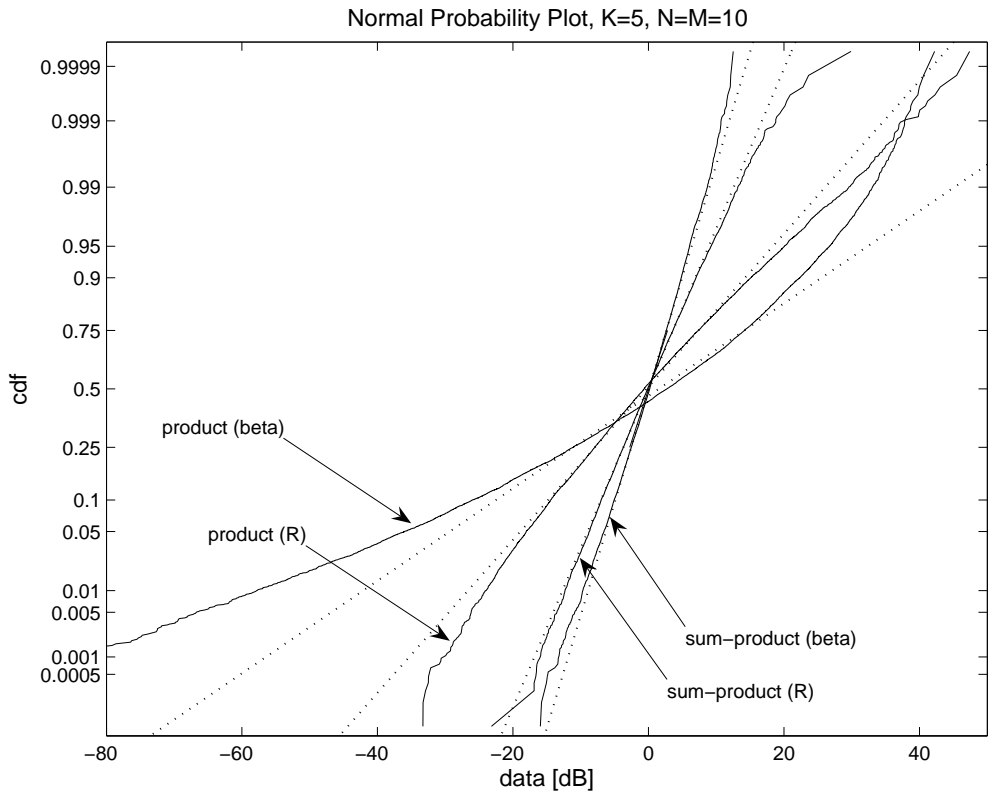

Fig. 7. Examples of empirical distributions of $10 \log_{10}(P)$ from Fig. 6 for $K = 5$. The straight dotted line displays the normal distribution. See Fig. 4 for $K = 40$.

simplicity that they are $N \times N$). With all tested distributions, the sum-product model gives a better fit for all values of $N$. Interestingly, with the sum-product model the fit improves as $N$ increases for the uniform and L distributions (for the R distribution no conclusions can be made about convergence for the examined range of $N$). Such convergence to log-normal distribution is somewhat surprising, since one would expect exactly the opposite since increasing the length of $\mathbf{a}$ and $\mathbf{b}$ implies increased averaging (summation) in the signal model. It appears that the increase in the size of the $N \times N$ coupling matrices balances the averaging effect.

The averaging effect from increasing $N$ can be seen in the standard deviation of the sum-product distribution, see Table III. The standard deviation of the distribution decreases sharply as $N$ increases. With the product model, $N$ has little effect on the standard deviation, which is in line with the earlier observation that number of rays affects only the distribution of the first

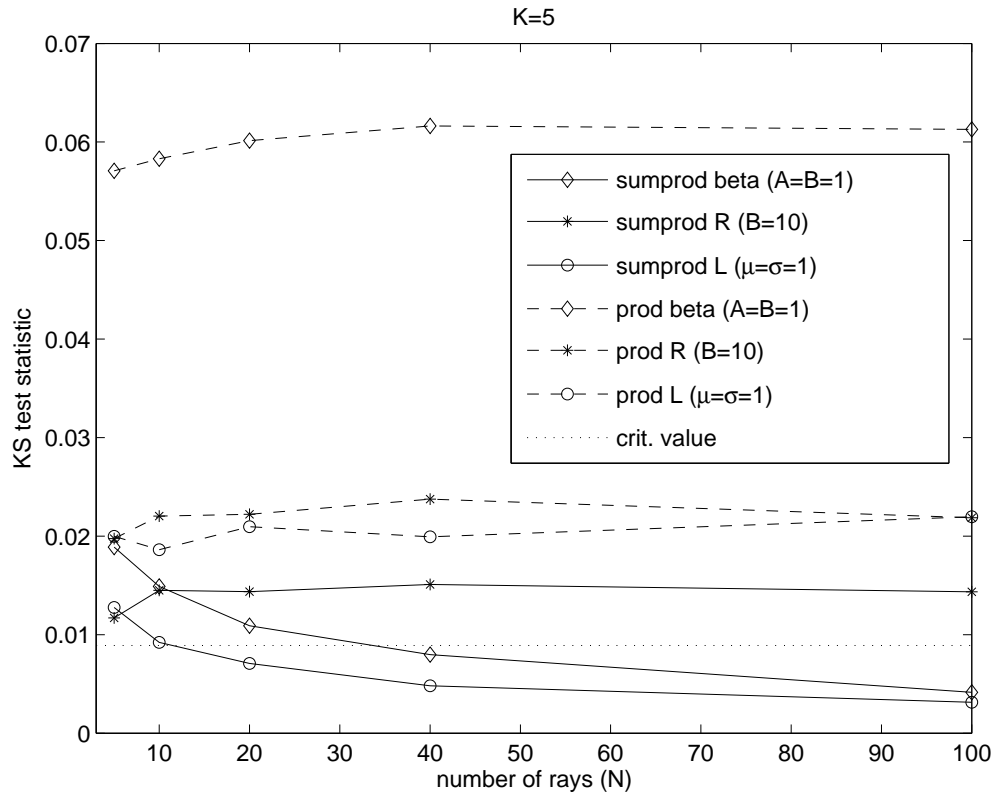

Fig. 8. Impact of number of rays, $N$, for different distributions. Small value of the test statistics indicates good fit. Critical value of the Kolmogorov-Smirnov test for 95% significance and $10^3$ samples is shown for reference.

factor in (8); this will not have much effect unless $K$ is very small.

### C. Impact of distributions

Fig. 9 illustrates the impact of distribution parameters on converge speed. Some remarks on the results:

- With all tested interaction distributions, including many others not shown in Fig. 9, exhibit convergence to the normal distribution. This implies that (6) converges to log-normal under very mild statistical assumptions on elements of the component matrices and vectors. As we are not aware of any limit theorems that would apply to the signal model (6), the requied conditions are not known.

- As is intuitively clear, the speed of convergence depends on the distribution of the interac-

TABLE III

STANDARD DEVIATIONS FOR FIG. 8

| model/pdf | std for $N = 5/10/20/40/100$ [dB] |
|-----------|-----------------------------------|
| sumprod/beta | 5.6/3.9/2.7/1.9/1.2 |
| sumprod/R | 7.6/5.6/4.0/3.0/1.9 |
| sumprod/L | 6.1/4.2/2.9/2.1/1.3 |
| prod/beta | 19.7/19.6/19.6/19.6/19.5 |
| prod/R | 11.7/11.4/11.2/11.0/10.9 |
| prod/L | 14.2/14.0/13.6/13.8/13.8 |

tions as well as its parameters.

• The fact that the convergence of the quadratic form (3) to a log-normal random is insensitive to the probability law of the individual entries of its component matrices implies that the sum-product model is capable of explaining the log-normality of the shadowing phenomenon for a wide variety of different channel interactions and radio propagation scenarios.

## VI. CONCLUSION

We have studied the applicability of a general sum-product signal model as a basis to explain the fluctuation of locally averaged received signal power in mobile communications, i.e., shadow fading. It was shown by simulations that as the number of multiplicative "layers of channel interactions" (coupling matrices) in the model increases, the distribution of the local average power tends to log-normal distribution, as observed in measurements. It was demonstrated that only a few layers (e.g. $3 - 5$) are needed to produce approximately log-normal power, whereas with the conventional model a product of more than $20$ random attenuations are needed to produce the same goodness-of-fit. Furthermore, the sum-product model does not make the implausible assumption that all multipaths experience the same cascade of random attenuations. The sum model, which addresses 'local' shadow fading, emerges as a special instance of the general model in the case, where the intermediate propagation process does not change and only the powers of the plane waves at the receiver end vary. The sum-product model, however, can explain log-normality of shadow fading in both local and global scale, and is therefore more general.

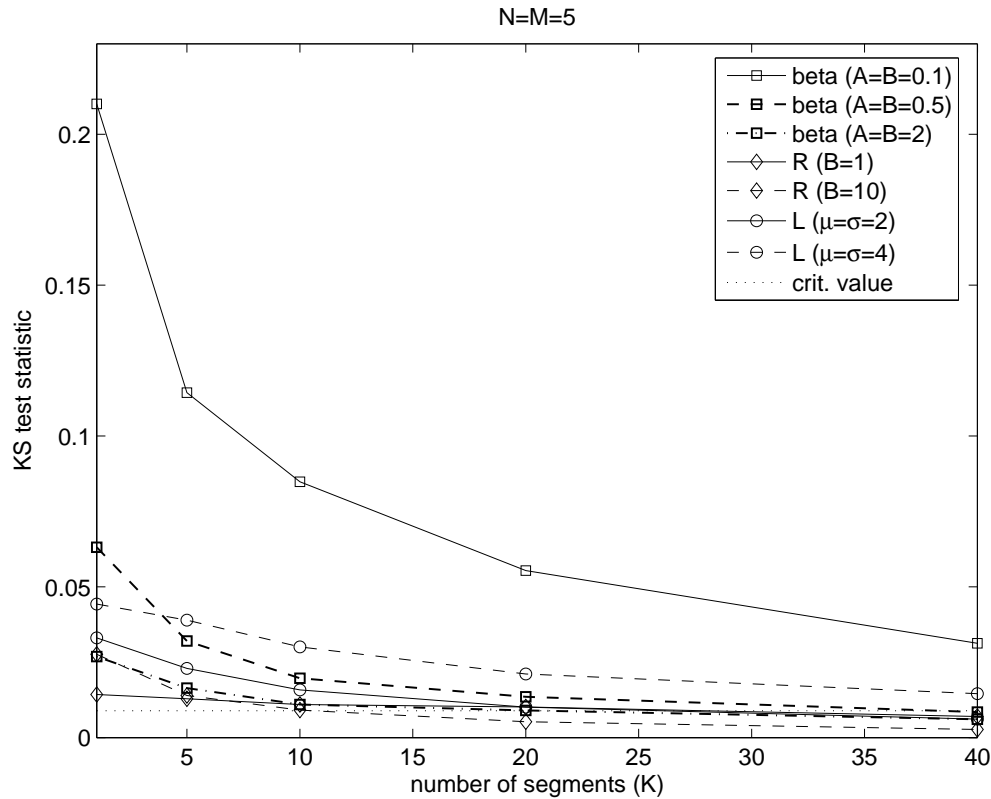

Fig. 9. Impact of distribution of interaction attenuations to $10\log_{10}(P)$ for $N = M = 5$. The straight dotted line displays the normal distribution. Small value of the test statistics indicates good fit. Critical value of the Kolmogorov-Smirnov test for $95\%$ significance and $10^3$ samples is shown for reference.

However, more work is needed to derive an analytical justification for the justification of the convergence of the sum-product model to a log-normal random variable.

## ACKNOWLEDGMENT

Comments from Vittorio Degli-Esposti and Nicolai Czink are gratefully acknowledged.